\documentclass[sigconf,authorversion]{acmart}

\usepackage{paralist}
\usepackage{listings}
\usepackage{xcolor}
\usepackage{setspace}
\usepackage{makecell}
\usepackage{multirow}
\usepackage{booktabs}

\newfloat{lstfloat}{t!}{lop}
\floatname{lstfloat}{Listing}

\usepackage{array}
\newcolumntype{L}[1]{>{\raggedright\let\newline\\\arraybackslash\hspace{0pt}}m{#1}}
\newcolumntype{C}[1]{>{\centering\let\newline\\\arraybackslash\hspace{0pt}}m{#1}}
\newcolumntype{R}[1]{>{\raggedleft\let\newline\\\arraybackslash\hspace{0pt}}m{#1}}
\newcolumntype{Z}[1]{>{\let\newline\\\arraybackslash\hspace{0pt}}m{#1}}
\newcolumntype{P}[1]{>{\centering\arraybackslash}p{#1}}

\AtBeginDocument{%
  }

\setcopyright{acmlicensed}
\copyrightyear{2018}
\acmYear{2018}
\acmDOI{XXXXXXX.XXXXXXX}

\begin{document}
\title[LAMeD: LLM-Generated Annotations for Memory Leak Detection]{LAMeD: LLM-generated Annotations for Memory Leak Detection}

\author{Ekaterina Shemetova}
\email{katyacyfra@gmail.com}
\orcid{0000-0002-1577-8347}
\affiliation{
  \institution{St. Petersburg University}%
  \city{St. Petersburg}
  \country{Russia}
}

\author{Ilya Shenbin}
\orcid{0000-0002-6778-225X}
\affiliation{%
  \institution{PDMI RAS}
  \city{St. Petersburg}
  \country{Russia}
}

\author{Ivan Smirnov}
\orcid{0009-0002-3690-4733}
\affiliation{%
  \institution{ITMO University}
  \city{St. Petersburg}
  \country{Russia}
}

\author{Anton Alekseev}
\orcid{0000-0001-6456-3329}
\affiliation{%
  \institution{PDMI RAS}
  \institution{St. Petersburg University}
  \city{St. Petersburg}
  \country{Russia}}
\affiliation{%
  \institution{KSTU n. a. I. Razzakov}
  \city{Bishkek}
  \country{Kyrgyzstan}
}

\author{Alexey Rukhovich}
\orcid{0000-0002-5701-1785}
\affiliation{%
  \institution{AI Foundation and Algorithm Lab}
  \city{Moscow}
  \country{Russia}
}

\author{Sergey Nikolenko}
\orcid{0000-0001-7787-2251}
\affiliation{%
  \institution{St. Petersburg Department of the Steklov Institute of Mathematics}
  \city{St. Petersburg}
  \country{Russia}
}

\author{Vadim Lomshakov}
\orcid{0000-0001-8991-9264}
\affiliation{%
  \institution{PDMI RAS}
  \city{St. Petersburg}
  \country{Russia}
}

\author{Irina Piontkovskaya}
\orcid{0009-0003-0299-5849}
\affiliation{%
  \institution{AI Foundation and Algorithm Lab}
  \city{Moscow}
  \country{Russia}
}







\renewcommand{\shortauthors}{Shemetova et al.}

\begin{abstract}
Static analysis tools are widely used to detect software bugs and vulnerabilities but often struggle with scalability and efficiency in complex codebases. Traditional approaches rely on manually crafted annotations---labeling functions as sources or sinks---to track data flows, e.g., ensuring that allocated memory is eventually freed, and code analysis tools such as \emph{CodeQL}, \emph{Infer}, or \emph{Cooddy} can use function specifications, but manual annotation is laborious and error-prone, especially for large or third-party libraries. We present LAMeD (LLM-generated Annotations for Memory leak Detection), a novel approach that leverages large language models (LLMs) to automatically generate function-specific annotations. When integrated with analyzers such as \emph{Cooddy}, LAMeD significantly improves memory leak detection and reduces path explosion. We also suggest directions for extending LAMeD to broader code analysis.
\end{abstract}



\maketitle




\section{Introduction}\label{sec:intro}

Memory leaks are a critical class of software defects that waste system resources, degrade performance, and can lead to system failures. Consequences can be severe and affect millions of users, including the gradual slowdown of web browsers or high-profile incidents like the 2012 AWS outage\footnote{\url{https://techcrunch.com/2012/10/27/amazon-web-services-outage-caused-by-memory-leak-and-failure-in-monitoring-alarm/}}, which websites \emph{Reddit} and \emph{Foursquare} among others, or the 2017 Bitcoin crash due to an alleged memory leak\footnote{\url{https://www.testbytes.net/blog/12-software-bugs-that-caused-epic-failure/}}. Vulnerabilities like OpenSSL's \emph{Heartbleed}~\cite{10.1145/2663716.2663755} show how improper memory management can lead to catastrophic outcomes.
%
%
Memory leaks are notoriously hard to detect, often requiring sophisticated techniques beyond manual inspection. Several automated approaches have been developed to aid in this process, including \emph{binary analysis}, which focuses on examining compiled code to identify abnormal memory usage patterns, \emph{source-code-based analysis}, which inspects source code and uses symbolic reasoning 
and data flow analysis to evaluate potential states without executing them, and \emph{dynamic program analysis}, which integrates runtime execution with symbolic methods. Despite their strengths, these methods have limitations when dealing with large codebases and complex memory management systems.

\begin{figure}[!t]
    \centering
    \includegraphics[width=0.99\linewidth]{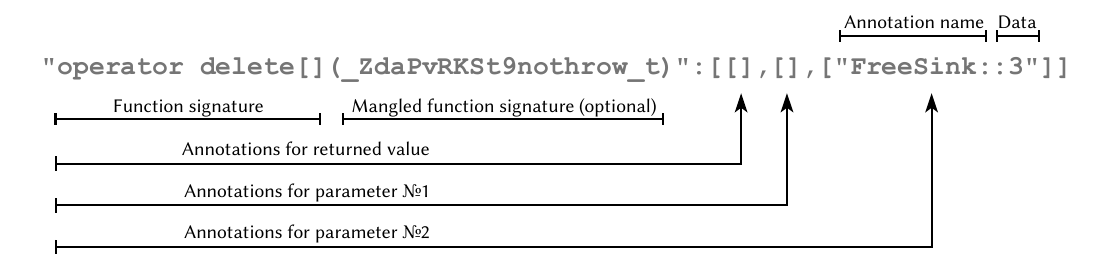}
    \caption{Annotating an ``\texttt{operator delete[](void*, std:: nothrow\_t const\&)}'' function (from \emph{Cooddy} docs; \texttt{FreeSink} means freeing a variable with a \texttt{free} function of ``type 3'').}
    \label{fig:syntax-cooddy-annotations}
\end{figure}

Static analysis methods are effective at uncovering subtle errors by systematically exploring various program paths, but they typically depend on rule-based annotations that designate functions as memory allocation sources or deallocation sinks, and are vulnerable to \emph{path explosion}, an exponential growth in execution paths in complex systems. Many static memory leak detectors, including \emph{CodeQL}, \emph{Semgrep}, and \emph{Cooddy}, rely on rules and specifications---hereafter called \emph{annotations}---to mark functions as allocation or deallocation points. For instance, \emph{CodeQL} enables custom allocation models (e.g., for OpenSSL functions such as \texttt{CRYPTO\_malloc} and \texttt{CRYPTO\_zalloc}) to refine its analysis. Similarly, \emph{Cooddy} uses structured annotations (as shown in Fig.~\ref{fig:syntax-cooddy-annotations}) to indicate whether memory is allocated or freed. Although some tools can derive these annotations automatically, doing so remains challenging due to the need for heuristics that vary across codebases. {For example, \emph{CodeQL} marks a function as allocating based on its name and signature, missing many functions. Kernel memory leak detector \emph{K-Meld}~\cite{emamdoost2021detecting} uses kernel-specific heuristics and context-aware rule mining to detect functions of interest.} Manual annotation by domain experts remains common despite its limited scalability and accuracy.

To address these problems, in this work we introduce \emph{LAMeD} (\textbf{L}LM-generated \textbf{A}nnotations for \textbf{Me}mory leak \textbf{D}etection), a novel approach that employs large language models (LLMs) to automatically generate tool-specific annotations for C/C++ functions, thereby enhancing memory leak detection, especially in complex, challenging codebases. LAMeD is designed to overcome key challenges in static analysis such as path explosion and the high overhead of manual annotation. Specifically, we address the following research questions:

\begin{itemize}
    \item[\textbf{RQ1}:] {How do LLM-generated annotations compare with manual and heuristic-based annotations?}
    \item[\textbf{RQ2}:] Can LLM-generated annotations improve the accuracy of real-life memory leak detection?
    \item[\textbf{RQ3}:] {How do LLM-generated annotations influence the number of warnings reported by the analyzer as compared to the ``no annotations'' setting?}
\end{itemize}

Our experimental evaluation shows that integrating LLM-ge\-ne\-rated annotations with static analysis tools increases real-life memory leak detections in complex codebases for both of the two considered static code analyzers. However, for some projects the number of the analyzers warnings significantly grows as a result. 

The rest of the paper is structured as follows: Section~\ref{sec:preliminaries} provides background on static analysis and LLMs, Section~\ref{sec:idea} outlines our proposed approach, Section~\ref{sec:demo_example} presents a motivating example, Section~\ref{sec:eval} shows our comprehensive evaluation study and discusses its results, Section~\ref{sec:related_work} reviews related research, and Section~\ref{sec:conclusion} concludes the paper and suggests directions for future work.


\section{Preliminaries}\label{sec:preliminaries}


In this section, we review the necessary background for this work, focusing on static program analysis for memory leak detection and using large language models (LLMs) for source code analysis.

\subsection{Static Program Analysis}





Static analysis encompasses a wide range of techniques that inspect source or object code without executing it. These methods are known to be effective in identifying deep-seated bugs, such as memory leaks, that often elude conventional testing methods. However, two key challenges remain:

\begin{itemize}
    \item \emph{scalability}: as codebases increase in size and complexity, the number of potential execution paths grows exponentially, and this path explosion makes it computationally expensive---or even infeasible---to examine every possible path;
    \item \emph{need for customization}: out-of-the-box static analysis tools come with built-in rules that often fail to capture the behaviors of a specific codebase; to improve accuracy, users have to supply custom annotations (e.g., designating certain functions as sources for data input or memory allocation and others as sinks for deallocation); this is critical for memory leak detection since standard functions such as \texttt{malloc} and \texttt{free} are often found only deep in the call stack, obscured by custom or external memory management routines, leading to key paths being overlooked and memory leaks missed.
\end{itemize}

\subsection{LLMs for Source Code Analysis}\label{ssec:preliminary_llm4code}

Large Language Models (LLMs) are Transformer-based~\cite{10.5555/3295222.3295349} architectures trained on massive datasets, including natural language and code. While originally developed for natural language processing, modern LLMs such as OpenAI's \emph{Codex}, Meta's \emph{CodeLlama}~\cite{roziere2023code}, DeepSeek-Coder~\cite{guo2024deepseek}, and Codestral\footnote{\url{https://mistral.ai/news/codestral/}} have shown exceptional capabilities in understanding and reasoning about code, solving tasks such as code generation, completion, and repair. They offer several important advantages for code analysis:
\begin{itemize}
    \item \emph{automated annotation}: LLMs can generate function-specific annotations that guide static analyzers, effectively reducing the number of paths that need to be explored and addressing the path explosion problem;
    \item \emph{tailored insights}: by understanding the context and relationships between functions, LLMs produce annotations that are more comprehensive and accurate than manual efforts;
    \item \emph{adaptability}: operating effectively in zero-shot or few-shot settings, LLMs can adapt to new codebases without extensive fine-tuning; their ability to process natural language comments and documentation further enables them to infer programmer intent, which is especially useful for legacy code with sparse documentation.
\end{itemize}


\section{Motivation and Idea}\label{sec:idea}

In this section, we present the core motivation behind integrating LLMs with {static analysis} for memory leak detection,  
illustrate the limitations of existing methods, and introduce our approach.

Static code analyzers typically rely on predefined \emph{rules} or ``hints'' called \emph{annotations} to identify key code elements---functions, variables, and data flows---that may indicate potential vulnerabilities. E.g., marking certain functions as \emph{sources} (e.g., those allocating memory via \texttt{malloc}) and \emph{sinks} (that perform deallocation via \texttt{free}) can guide analyzers in detecting memory leaks, increasing the number of true positive detections by making the search process more precise and automatically excluding irrelevant paths. In mission-critical projects, annotations are typically written manually by specialists. However, manual annotation is both laborious and error-prone, and does not scale well to ever-evolving codebases in real-world projects, where custom memory management functions are common. These limitations motivate a more automated approach.



Inspired by advances in natural language processing, where human annotation is increasingly supplemented by generative models~\cite{nasution2024chatgptlabel, alizadeh2024opensourcellmstextannotation,kim2024megannohumanllmcollaborativeannotation,lee2023rlaif}, we propose to leverage LLMs trained on public source code (see Section~\ref{sec:related_work} and surveys~\cite{zhang2024unifyingperspectivesnlpsoftware,awesome-code-llm}) to generate custom, analyzer-specific annotations for functions in the codebase. Our approach is based on the premise that LLMs trained on vast code repositories can reliably identify code segments responsible for memory allocation and deallocation even in the presence of non-standard structures or unconventional naming conventions. Unlike traditional formal methods, LLMs can use variable names, function names, and inline comments in their reasoning, yielding flexible and context-aware annotations. In this work, we present a~proof of~concept for~this approach by~focusing on~prompt-based annotation generation without model fine-tuning. Our results, presented in Section~\ref{sec:eval}, show encouraging results.
The proposed approach can be structured as the pipeline shown in Fig.~\ref{fig:memleak_pipeline}:

\begin{figure*}[!t]
    \centering    
    \includegraphics[width=0.85\linewidth]{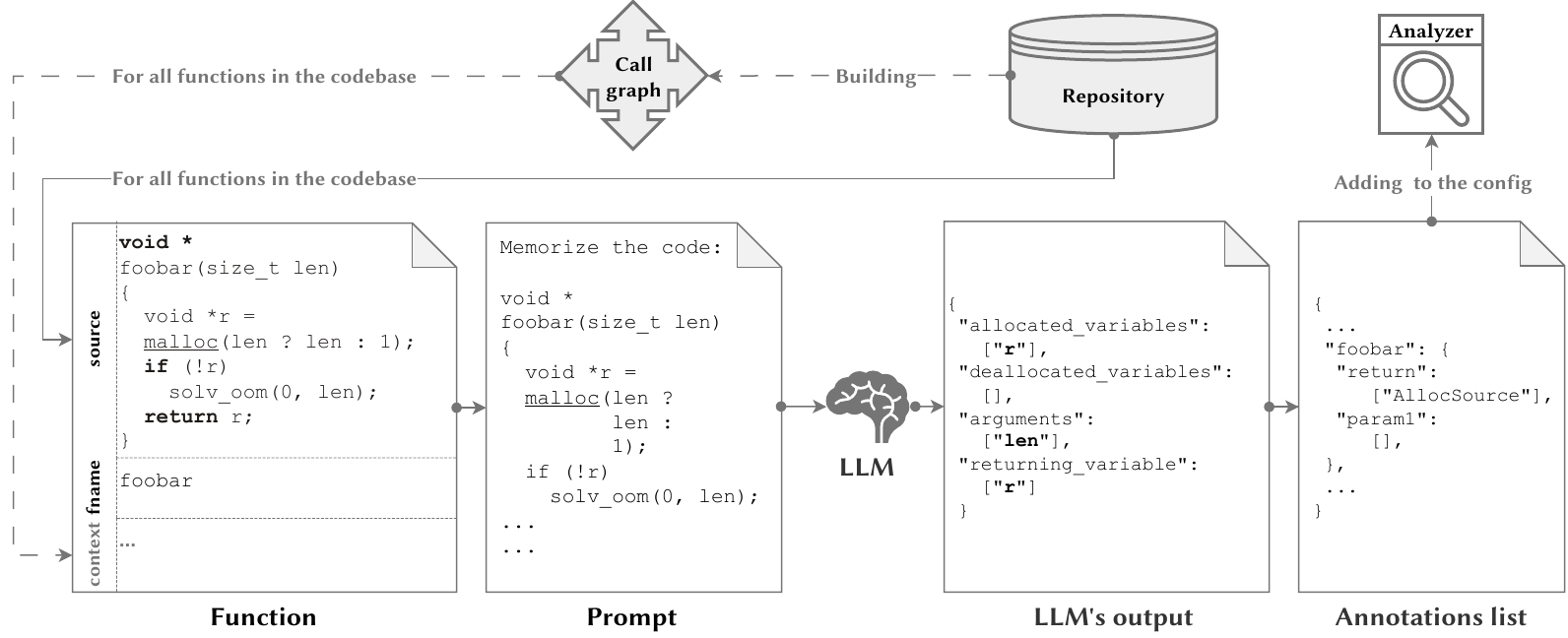}
    \caption{The proposed LLM-enhanced memory leak detection pipeline.}
    \label{fig:memleak_pipeline}
\end{figure*}

\begin{enumerate}[(1)]
    \item extract the call graph and source code of all functions (and macros, if~relevant) from~the~codebase;
    \item 
    extract context information from~neighbors in~the~call graph;
    \item prompt the LLM with the source code of each function and its call graph context to identify variables and arguments involved in memory [de]allocation, in JSON format~\cite{pezoa2016foundations};
    \item convert LLM outputs into function-specific annotations (the annotation set for some functions may be, and often is, empty);
    \item provide generated annotations to the memory leak detector (either in its configuration or in the code itself);
    \item execute the analysis with annotations.
\end{enumerate}

Since querying an LLM incurs a significant cost, the annotation generation is less frequent---e.g., after major code changes---and its results are reused across many analysis runs. Similarly, call graph construction is done incrementally to reduce overhead; see Section~\ref{sec:demo_example} for an~illustrative example.
We begin by preprocessing the code by splitting it into individual functions and constructing a call graph to provide contextual information for each function. Using the \emph{Joern} engine~\cite{joern_io_Joern_The_Bug_2024}, we extract caller-callee relationships, function code, and other pertinent data to enrich the LLM prompts.
For each function, we perform a two-step LLM query:
\begin{itemize}
    \item \emph{initial query}: we prompt the LLM (see Listing~\ref{lst:prompt1}) with the function’s source code, optionally supplemented by the source of its callees, to obtain basic details such as argument names, the return variable, and variables holding pointers to newly allocated or freed memory;
    \item \emph{post-processing}: for functions flagged as allocating, we further validate that the memory is allocated to a new variable (and not merely assigned to an existing object or structure) by prompting the LLM again (see Listing~\ref{lst:prompt3}); this prompt includes the source code of the function, its callees, and the relevant structure and argument names.
\end{itemize}
Finally, the LLM's output is transformed into a JSON format compatible with the static analysis tool.

\begin{lstfloat}
\begin{lstlisting}[basicstyle=\ttfamily\footnotesize,caption={
%The prompt template, where 
Prompt template for annotation generation.
\texttt{\{code\}} is the source code of the function and \texttt{\{func\_name\}} is its name (without arguments).},columns=fullflexible,keepspaces=true,label={lst:prompt1},frame=single]
You are a C developer. Your task is to answer the following 
questions about a code snippet.
Which variables contain pointers to the memory allocated in
function {func_name}? Put the answer in the "allocated_variables"
field.
Which variables contain pointers to the memory deallocated in
function {func_name}? Put the answer in the "deallocated_variables"
field. 
Return the final answer as a short JSON object.

# code
{code}
\end{lstlisting}

\begin{lstlisting}[basicstyle=\ttfamily\footnotesize,caption={
%The prompt template, where 
Prompt template for an additional check;
\texttt{\{source\}} is the source code of the function and its callees, \texttt{\{func\_name\}}, name of the function, \texttt{\{structure\}}, name of the structure passed by reference, \texttt{\{variable\_name\}}, name of the argument.},columns=fullflexible,keepspaces=true,label={lst:prompt3},frame=single]
You are a C developer. Your task is to answer the following 
question about a code snippet. Does the returned value of the
function {func_name} point to the part of {structure} structure 
which is passed as an argument {variable_name}? 

Give a detailed answer in JSON format without any comments.

{source}
\end{lstlisting}
\end{lstfloat}

\section{Illustrative Example}\label{sec:demo_example}

A memory leak occurs when a program allocates memory that is never subsequently released, eventually leading to resource depletion. Consider the example in Listing~\ref{lst:demo_poc}, which demonstrates a real memory leak that was fixed in the \texttt{libsolv} library\footnote{\scriptsize\url{https://github.com/openSUSE/libsolv/commit/98a75959e13699e2ef35b0b011a88a6d224f227e\#diff-06a5bd14eb15124f5d8d2e346f7226cb8b4ec47b8a205a77fb51c276f227781eL1900}}. In this case, memory is allocated to the variable \texttt{leadsigchksum} by the function \texttt{solv\_chksum\_create}, but the corresponding deallocation is omitted at line 1900, resulting in a leak.

The typical strategy for detecting such leaks involves tracing the data flow from the allocation point to the program's exit or deallocation routines. Standard analyzers often identify allocation functions such as \texttt{malloc} or \texttt{realloc}. However, when custom allocation functions such as \texttt{solv\_chksum\_create} are used, the analyzer must inspect not only the function itself but also its callees to confirm that the allocated memory is eventually freed. This task becomes even more challenging when the function is part of a third-party library that is not readily available for analysis.

To mitigate these challenges, the analyzer's configuration can be augmented with additional information about the custom functions. For instance, annotating the \texttt{solv\_chksum\_create} function as an allocating function (see Listing~\ref{lst:demo_anno}) guides the \emph{Cooddy} analyzer to check that the memory returned by this function is properly deallocated. This targeted annotation compensates for the limitations of generic heuristics, ensuring that critical memory leaks are not overlooked even in complex codebases.

\begin{lstfloat}
\begin{lstlisting}[firstnumber=1890,basicstyle=\ttfamily\footnotesize,caption={Code demonstrating a memory leak vulnerability in the \texttt{libsolv} library.},columns=flexible,keepspaces=true,label={lst:demo_poc}, numbers=left,numbersep=5pt,framexleftmargin=0mm,language=C++,frame=single]
...
  /* process lead */
if (chksumtype)
  chksumh = solv_chksum_create(chksumtype);
if ((flags & RPM_ADD_WITH_LEADSIGID) != 0)
  leadsigchksumh = solv_chksum_create(REPOKEY_TYPE_MD5);
if (fread(lead, 96 + 16, 1, fp) != 1 || getu32(lead) != 0xedabeedb)
{
  pool_error(pool, -1, "%s: not a rpm", rpm);
  fclose(fp);
  return 0;
}
...
\end{lstlisting}
\setstretch{0.85}
\begin{lstlisting}[firstnumber=1,basicstyle=\ttfamily\footnotesize,caption={A sample \emph{Cooddy} annotation for the \texttt{solv\_chksum\_create} function from \texttt{libsolv}, which hints that the returned value should be deallocated.},columns=flexible,keepspaces=true,label={lst:demo_anno}, numbers=left,numbersep=5pt,framexleftmargin=0mm,language=C++,frame=single]
  "solv_chksum_create(solv_chksum_create)": [
    [
      "AllocSource::1"
    ],
    []
  ]
\end{lstlisting}
\end{lstfloat}

\section{Evaluation}\label{sec:eval}


In this section, we present our experimental setup and show how LLM-based annotation generation enhances memory leak detection by static analysis tools. We measure both the accuracy of the annotations in identifying memory leaks and their impact on the overall performance of the analyzers. We compare the LLM-enhanced versions of \emph{CodeQL} and \emph{Cooddy} against their baseline configurations using two datasets: a manually annotated cJSON library and a real-life memory leak dataset drawn from several C projects.

\subsection{Experimental Setup}\label{ssec:eval_experimental_setup}

\subsubsection{Static Analysis Tools}

In our experiments, we used three static analysis tools: \emph{CodeQL}\footnote{\url{https://codeql.github.com/}}, \emph{Cooddy}\footnote{An open source version at \url{https://github.com/program-analysis-team/cooddy}}, and \emph{Infer}.

\emph{CodeQL} is an industry-leading semantic code analysis engine with a wide array of built-in rules and queries for code analysis, as well as facilities for creating custom rules. In our experiments with \emph{CodeQL}, we extended the set of \emph{CodeQL}'s C/C++ models\footnote{\url{https://github.com/github/codeql/tree/main/cpp/ql/lib/ext}} by incorporating custom memory models derived from our annotated function descriptions. For leak detection, we used the built-in query sets \texttt{MemoryNeverFreed.ql}\footnote{\url{https://github.com/github/codeql/blob/main/cpp/ql/src/Critical/MemoryNeverFreed.ql}} and \texttt{MemoryMayNotBeFreed.ql}\footnote{\url{https://github.com/github/codeql/blob/main/cpp/ql/src/Critical/MemoryMayNotBeFreed.ql}}, which operate in conjunction with our custom (generated) and built-in C/C++ models. Additionally, \emph{CodeQL} includes a heuristic named \emph{HeuristicAllocationFunctionByName}\footnote{\url{https://github.com/github/codeql/blob/631ccdf380d12ae0d4720d2e9e1c32be6bca1693/cpp/ql/lib/semmle/code/cpp/models/implementations/Allocation.qll\#L365}} for recognizing additional allocation functions, which we also executed separately.

\emph{Cooddy} is a less widely known tool\footnote{See documentation at \url{https://github.com/program-analysis-team/cooddy/tree/master/docs} and in the work~\cite{gerasimov2022case}.}, an extensible static code analysis engine designed for C/C++/Java. \emph{Cooddy} can perform both data flow analysis (DFA) and static symbolic execution (SSE). It provides a set of checkers for various common faults and allows to add custom checkers. For memory leak detection, we used the built-in \texttt{MemoryAndResourceLeakChecker}\footnote{\url{https://github.com/program-analysis-team/cooddy/blob/master/docs/MemoryAndResourceLeakChecker.md}}, configuring it with function annotations \texttt{AllocSource} for allocation functions and \texttt{FreeSink} for deallocation functions\footnote{\url{https://github.com/program-analysis-team/cooddy/blob/master/docs/Annotations.md}}.


For a comparative analysis of LLM-enhanced and ``vanilla'' versions of \emph{Cooddy} and \emph{CodeQL}, we conduct experiments using the widely adopted open-source static analyzer, \emph{Infer}, which targets a broad range of potential faults across a number of programming languages (see Section~\ref{sec:sourcecodeanalysis} for more references). 
We used \emph{Infer}'s \emph{Pulse} analyzer, focusing only on detections labeled as \texttt{MEMORY\_LEAK\_C}. For each project, we supplied a \texttt{compile\_commands.json} file, automatically generated via CMake or Bear~\cite{Bear} depending on the project's build configuration, and executed the ``\texttt{infer run}'' command, indicating that only \emph{Pulse} analyzer is to be used and providing the path to the compilation database. 

\subsubsection{Data}\label{ssec:data}


Since we use the LLM in a zero-shot setting, we only describe the datasets we have used for evaluation. Issues that are our main focus in this study are hard to {find and} reproduce, and
here we present our attempt to overcome this problem and evaluate the proposed approach from different perspectives. We construct datasets that would allow to study several aspects of LLM-enhanced static analysis targeted at memory leak detection.

{\paragraph{Manually annotated cJSON library.} Five authors of this paper labeled $30+$ methods of the open source cJSON library each with indicators whether the function's arguments or returned values should or should not be annotated as allocating or deallocating memory.}
Note that annotation guidelines cannot be complete (otherwise this task would be easy to solve via formal rule-based methods); annotators were instructed to choose functions that are used in the project as constructor/destructor-like functions and carefully analyze their usage: if a function allocates memory that should be deallocated separately, then the function is labeled as an allocating function.
%

{Initial manual annotations were reviewed, and especially ambiguous cases were resolved by an expert annotator (one of the authors). The resulting dataset corresponds to the latest version of the \emph{\href{https://github.com/DaveGamble/cJSON}{cJSON}} repository's \texttt{master} branch,\footnote{\scriptsize\url{https://github.com/DaveGamble/cJSON/tree/12c4bf1986c288950a3d06da757109a6aa1ece38}.} and comprises 152 functions, each optionally annotated with the tags \texttt{AllocSource} or \texttt{FreeSink}. None of the functions have both tags, 44 functions were marked as performing memory allocation (\texttt{AllocSource}), 11 as performing deallocation (\texttt{FreeSink}), and the remaining 97 functions were left unlabeled.}
{The purpose of this annotation scheme is to check whether LAMeD can find all allocating/deallocating functions. Also, it is important not to mark redundant functions since this may lead to a lot of false positives in the analysis. We use this manually labeled dataset to compare different LLM models and context settings. The best configuration in terms of precision/recall was then used to generate annotations for static code analysis tools on our real-life memory leak dataset described below.}

{\paragraph{Real-life memory leak dataset.} To evaluate the approach on real-world memory leaks, we selected several C projects from the \emph{DiverseVul}~\cite{chen2023diversevul} vulnerabilities dataset based on two criteria: easy compilation in modern environments (without complex setups involving outdated dependencies) and documented resolved memory leak issues available on \emph{GitHub}. For each selected case, we have manually chosen memory leaks (``target bugs'') that have been reported, confirmed, and fixed by developers. Each memory leak is associated with its own ``fixing'' commit, and the repository state corresponding to the parent commit was included to the dataset.} 
{Project-specific statistics are summarized in Table~\ref{tab:irl_dataset}. The table shows the average number of functions per repository (computed with \emph{Joern}), the number of target bugs for each project (one bug corresponds to one version of the repository), and results of the analyzers (\emph{CodeQL}, \emph{Cooddy}, \emph{Infer}) without any customization on the dataset. While reliably quantifying false-positive rates of memory leak detectors is challenging in such a scenario---additional, unrelated memory issues may exist---this dataset allows to verify if a detector can identify known memory leaks. Moreover, real-world memory-related bugs typically represent more challenging detection scenarios.}

\begin{table}[!t]
    \caption{Real life memory leaks dataset and analyzer results.}
    \label{tab:irl_dataset}
    \setlength{\tabcolsep}{3pt}
    \begin{tabular}{lccccc}
        \toprule
        \multirow{2}{*}{\textbf{Project}}  & \multicolumn{1}{l}{\multirow{2}{*}{\textbf{\begin{tabular}[c]{@{}l@{}}\# of  \\ functions\end{tabular}}}}  & \multicolumn{1}{l}{\multirow{2}{*}{\textbf{\begin{tabular}[c]{@{}l@{}}\# of  target\\ bugs\end{tabular}}}}  & \multicolumn{3}{c}{\textbf{\begin{tabular}[c]{@{}c@{}}\# of target bugs  found\end{tabular}}}  \\ 
         & \multicolumn{1}{l}{}  & \multicolumn{1}{l}{}  & \multicolumn{1}{l}{\textbf{CodeQL}}  & \multicolumn{1}{l}{\textbf{Cooddy}}  & \multicolumn{1}{l}{\textbf{Infer}}  \\ 
        \midrule
        cjson       & 152   & 6  & {0}  & {0}  & 0  \\
        curl        & 2250  & 14 & {2}  & {1}  & 0  \\
        libsolv     & 1513  & 5  & {0}  & {1}  & 1  \\
        libtiff     & 1103  & 9  & {2}  & {3}  & 1  \\
        libxml2     & 2695  & 4  & {0}  & {0}  & 0  \\
        rabbitmq-c  & 274   & 2  & {1}  & {0}  & 0  \\
        libssh2     & 473   & 3  & {0}  & {0}  & 0  \\ 
        \midrule
        \textbf{Total}  & \textbf{8460}  & \textbf{43}  & {\textbf{5}}  &{\textbf{5}}  & \textbf{2}  \\ 
        \bottomrule
    \end{tabular}
\end{table}

\subsection{Annotation Quality}\label{ssec:eval_anno_quality}

\subsubsection{cJSON Results}


To compare LLM-based annotations, we used the manually annotated cJSON library (Section~\ref{ssec:data}). We tested three LLMs: Codestral~\cite{Codestral-22B-v0.1}, Qwen2.5-Coder-32B~\cite{hui2024qwen2} and DeepSeekR1-70B~\cite{guo2025deepseek}\footnote{Llama distillation of the R1 model available at \url{https://huggingface.co/deepseek-ai/DeepSeek-R1-Distill-Llama-70B}}. 
In our experiments, these models provided responses that were properly structured (as JSON) and were consistent across different queries; meanwhile, the content quality of those responses was also significantly ahead of their alternatives. 
Apart from that, we prioritized working with models that could be deployed and queried offline in a local network, since processing proprietary code via an external service may not be feasible for projects with security restrictions.

On this step, we evaluate of the impact of additional context and filtering on the LLMs' performance in the task of annotation generation. The metrics were calculated based on the manual labeling, and the resulting comparison is shown in Table~\ref{tab:model_performance}. We count as a true positive (TP) a function that was annotated and correctly marked in terms of whether it allocates or deallocates. A function is considered to be a false positive (FP) if it was annotated with a different label. The false negative cases (FN) are functions that were not annotated with LLM (had an empty label set) but were annotated in the ground truth dataset. We computed the Precision and Recall metrics as follows: \begin{equation}\mathrm{Precision} = \frac{\mathrm{TP}}{\mathrm{TP}+\mathrm{FP}},\quad
   \mathrm{Recall} = \frac{\mathrm{TP}}{\mathrm{TP}+\mathrm{FN}}.\end{equation}

{In our comparison, we considered different configurations: functions provided in prompts enriched with called functions (callees) as context and without; moreover, we tested how post-filtering affects the results.}
{Based on the comparison, models provide different total numbers of generated annotations; \emph{DeepSeek-R1-70B} tends to generate more than others. Thus, it expectedly delivers the highest recall value \emph{0.646} and the lowest precision value \emph{0.574}. Post-filtering allows to decrease the total number of annotated functions by about 20-25\% and noticeably increase precision for every run. Besides, additional context worked differently for the models, and it was only with \emph{Qwen2.5-Coder-32B} that adding the context increased the precision.}

The model with the largest number of parameters, \emph{DeepSeek-R1-70B}, did not provide the better results in comparison, which can be connected with the code-related training of two other models. The results for \emph{Codestral} and \emph{Qwen2.5-Coder} are close in average for the task, but \emph{Codestral} has fewer parameters (22B) and is thus faster during inference, which may be important for large-scale use cases of LLM-based annotations. For this reason, we chose to perform real-life memory leak detection quality estimation with the \emph{Codestral} model with post-filtering.

\begin{table}[!t]
    \caption{LLM-based annotations on cJSON: true positives (TP), false positives (FP), false negatives (FN), precision and recall; CE~--- context-enriched, PF~--- post-filtered, \textbf{\#}~--- total annotations generated.}
    \label{tab:model_performance}
    \setlength{\tabcolsep}{3pt}
    \begin{tabular}{lcccccccc}
        \toprule
        \textbf{Model}  & \textbf{CE}  & \textbf{PF}  & \textbf{TP}  & \textbf{FP}  & \textbf{FN}  & \textbf{Prec}  & \textbf{Rec}  & \textbf{\#}  \\ 
        \midrule
        DeepSeek-R1-70B    & ---  & ---  & 31  & 23  & 17  & 0.574  & \textbf{0.646}  & 54  \\
            & ---  & \checkmark  & 30  & 12  & 18  & 0.714  & 0.625  & 42  \\
            & \checkmark & ---  & 29  & 23  & 18  & 0.558  & 0.617  & 52  \\
            & \checkmark & \checkmark  & 28  & 13  & 19  & 0.683  & 0.596  & 41  \\
        \midrule
        Codestral          & ---  & ---  & 28  & 12  & 20  & 0.7  & 0.583  & 40  \\
                  & ---  & \checkmark  & 28  & 2  & 20  & \textbf{0.933}  & 0.583  & 31  \\
                  & \checkmark & ---  & 23  & 14  & 24  & 0.622  & 0.489  & 37  \\
                  & \checkmark & \checkmark  & 22  & 6  & 25  & 0.786  & 0.468  & 28  \\
        \midrule
        Qwen2.5-Coder-32B  & ---  & ---  & 31  & 12  & 16  & 0.721  & 0.66  & 43  \\
          & ---  & \checkmark  & 29  & 3  & 18  & 0.906  & 0.617  & 32  \\
          & \checkmark & ---  & 28  & 10  & 20  & 0.737  & 0.583  & 38  \\
          & \checkmark & \checkmark  & 27  & 2  & 21  & 0.931  & 0.562  & 29  \\
        \bottomrule
    \end{tabular}
\end{table}

\subsubsection{CodeQL heuristic vs. LAMeD annotations}

CodeQL has its own heuristic \emph{HeuristicAllocationFunctionByName}\footnote{\url{https://github.com/github/codeql/blob/631ccdf380d12ae0d4720d2e9e1c32be6bca1693/cpp/ql/lib/semmle/code/cpp/models/implementations/Allocation.qll\#L365}} for recognizing additional allocation functions. According to the documentation, it requires that
\begin{inparaenum}[(1)]
    \item the word \texttt{alloc} appears in the function name,
    \item the function returns a pointer type, and
    \item there is be a unique parameter of unsigned integer type.
\end{inparaenum}

We have generated annotations with \emph{CodeQL} for projects from our real-life dataset and checked the intersection sizes with LAMeD-generated annotations. The resulting comparison is shown in Table~\ref{tab:CodeQL_anno}. The second and third column show the average number of generated annotations per project by LAMeD and \emph{CodeQL} respectively, the column ``$A_{\mathrm{LAMeD}} \cap A_{\mathrm{CodeQL}}$'' shows the average number of common annotations in LAMeD and \emph{CodeQL} generated sets. The number of LAMeD-generated annotations significantly exceeds the number of \emph{CodeQL} heuristic annotations on all projects since \emph{CodeQL} uses only one heuristic {(such as function naming and signature)} while LAMeD can consider a large variety of cases. 
For most projects, LAMeD was able to find all allocation functions that were found by \emph{CodeQL}, but in the case of \texttt{cJSON} and \texttt{libxml2}, LAMeD was not able to recognize some ``low-level'' allocation functions such as \texttt{cJSON\_malloc} and \texttt{myMallocFunc}, {seemingly due to the fact that those functions do not directly contain variables that hold pointers to allocated memory. At the same time, \emph{CodeQL} annotated them correctly based on the name and signature.}

\begin{table}[!t]
    \caption{CodeQL vs. LAMeD annotation sets.}
    \label{tab:CodeQL_anno}
    \begin{tabular}{lccc}
        \toprule
        \textbf{Project}  & \multicolumn{1}{l}{\textbf{LAMeD}}  & \multicolumn{1}{l}{\textbf{CodeQL}}  & \multicolumn{1}{l}{\textbf{$A_{\mathrm{LAMeD}} \cap A_{\mathrm{CodeQL}}$}}  \\ 
        \midrule
        cjson       & 24  & 1  & 0  \\
        curl        & 117  & 1  & 1  \\
        libsolv     & 88  & 2  & 2  \\
        libtiff     & 18  & 0  & 0  \\
        libxml2     & 91  & 7  & 0  \\
        rabbitmq-c  & 10  & 1  & 1  \\
        libssh2     & 27  & 3  & 1  \\
        \bottomrule
    \end{tabular}
\end{table}

\subsection{Real-Life Memory Leak Detection Quality}\label{ssec:eval_leak_detection}

To check the impact of generated annotations on real memory leak detection by static code analyzers, we consider two important characteristics: the \emph{number of target bugs found} and the \emph{number of analyzer warnings} (with and without annotations). 

We have generated annotations for each repository in our real-life bug dataset; then, \emph{CodeQL} and \emph{Cooddy} were configured with generated annotations and run separately on each repository. Then we manually checked reports considering target files containing the target bug to confirm that the analyzer has found the bug. \emph{CodeQL} uses its own annotations, generated by \emph{HeuristicAllocationFunctionByName} in both (no annotation/annotation settings). Also, \emph{CodeQL} custom memory models use allocation functions in the assumption that the memory is always allocated to the returned value, so the LAMeD-generated annotation without this property were removed from consideration. \emph{Cooddy} allows fine-grained configuration considering variables to which memory is allocated to, so we used the full set of annotations with it. 

The results of \emph{CodeQL} and \emph{Cooddy} are shown in
Table~\ref{tab:irl}.
LAMeD-generated annotations help increase the number of found bugs in different projects for both \emph{Cooddy} and \emph{CodeQL}. Interestingly, no bug was found in \texttt{libxml2} despite the large number of annotations. The number of analyzer warnings with annotations increased significantly compared to the  ``without annotations'' setting for both analyzers. It is easy to see a direct relationship between the number of annotations and number of warnings. Note also that some sets of annotations are more effective than others, e.g., \texttt{cjson}, \texttt{libtiff}, \texttt{libssh2} annotations do not lead to a huge number of warnings but help find more bugs compared to the ``no annotations'' settings.

\begin{table}[!t]
    \caption{\emph{CodeQL} and \emph{Cooddy} memory real leak detection results with and without LAMeD-generated annotations.}
    \label{tab:irl}
    \begin{tabular}{lC{.12\linewidth}cccc}\toprule
        \multirow{2}{*}{\textbf{Project}}  & 
        \textbf{Avg. annot.}  &  \multicolumn{2}{C{.2\linewidth}}{\textbf{Target bugs found}} & 
        \multicolumn{2}{C{.25\linewidth}}{\textbf{Avg. analyzer warnings}}  \\
               & {}  & {\textbf{no ann.}}  & {\textbf{ann.}}  & {\textbf{no ann.}}  & \multicolumn{1}{l}{\textbf{ann.}}  \\\midrule
        \multicolumn{6}{c}{\textbf{\emph{CodeQL}}}\\\midrule
        cjson  & 24  & {0}  & 1  & {0}  & 29  \\
        curl  & 117  & {2}  & 3  & {13}  & 222  \\
        libsolv  & 88  & {0}  & 1  & {28}  & 158  \\
        libtiff  & 18  & {2}  & 2  & {65}  & 83  \\
        libxml2  & 91  & {0}  & 0  & {5}  & 118  \\
        rabbitmq-c  & 10  & {1}  & 1  & {13}  & 8  \\
        libssh2  & 27  & {0}  & 2  & {15}  & 35  \\ 
        \textbf{Total}  & \textbf{375}  & {\textbf{5}}  & \textbf{10}  & {\textbf{139}}  & \textbf{653}  \\ 
        \midrule
        \multicolumn{6}{c}{\textbf{\emph{Cooddy}}}\\\midrule
        cjson  & 24  & {0}  & 3  & {0}  & 9  \\
        curl  & 137  & {1}  & 2  & {47}  & 102  \\
        libsolv  & 103  & {1}  & 1  & {13}  & 115  \\
        libtiff  & 24  & {3}  & 4  & {13}  & 26  \\
        libxml2  & 120  & {0}  & 0  & {8}  & 114  \\
        rabbitmq-c  & 10  & {0}  & 0  & {1}  & 4  \\
        libssh2  & 53  & {0}  & 0  & {4}  & 21  \\ 
        \textbf{Total}  & \textbf{471}  & {\textbf{5}}  & \textbf{10}  & {\textbf{86}}  & \textbf{391}  \\ 
        \bottomrule
    \end{tabular}
\end{table}

\subsection{Discussion}\label{ssec:eval_discussion}

Combining the semantic code understanding of LLMs with reliability of formal code analysis is an important and promising area of the research. Usually, formal methods are used to check the quality of LLM-generated code; our LAMeD approach works in the opposite direction, utilizing LLMs to improve the coverage and efficiency of a formal code analyzer.

We chose \emph{Codestral} as a backbone model for LAMeD based on annotated cJSON library functions and carried out a number of experiments on a real-life dataset as shown above. {Few-shot prompting allows to increase the quality of LLM predictions on the target dataset. In this work, we focused on zero-shot prompts to compare results across different repositories without the impact of examples from specific ones.} We have arrived at the following conclusions.

First, we wanted to find out whether LLM prompting can be used as an aid or a replacement for a security specialist annotating sources and sinks for memory allocation-related processes in code. According to the manually annotated cJSON library, we can conclude that LLMs provide adequate precision and allow for making annotations more efficient since expert help will be needed less.

Based on a comparison of LLM-generated and an analyzer's heuristic-based annotations, such as \emph{CodeQL}, we can answer \textbf{RQ1}: we have found that LLMs allow to annotate more complex functions based on the context such as \texttt{solv\_chksum\_create}, while they can ignore allocation functions such as \texttt{cJSON\_malloc}, which can be labeled by simple name and signature-based heuristics.

As for the impact of the annotations on the memory leak detection quality (\textbf{RQ2}), 
the total number of detected bugs not previously found by the tool has increased on average. Given that our approach is arguably straightforward, this inspires optimism. 

It is important to note, however, that the extensive number of annotated function marked by LLM may increase the number of found bugs but the price to pay for that is high: the number of analyzer warnings can increase drastically (\textbf{RQ3}). 
In practice, it can be a problem for large codebases: modern static analyzers suffer from a high rate of false positives by default without any specific annotations/configuration. The growth in the number of warnings using annotations can be explained by 
\begin{inparaenum}[(1)]
    \item false positive allocation functions in the annotation set, when the analyzer starts to expect deallocation which is really not required; moreover, every single call of such function yields a false positive warning;
    \item the principle of operation of the analyzers themselves---even for perfect annotations false positives are possible, for example when the same function is used both for separate memory allocation (creates some object that should be destructed by its deallocation function) and for allocation as a part of more complicated structure (e.g., the allocated object is added to a queue-like object, then deallocated together with it by queue destructor).
\end{inparaenum}
In the first case, object allocation requires its own deallocation, while in the second case it does not. Importantly, in cases with a large number
of annotated functions an analyzer's performance may get worse depending on the analyzer architecture. However, we believe that there is an effective tradeoff between the number and completeness of the annotated functions and the memory leak detection effectiveness.





\section{Related Work}\label{sec:related_work}

In this section, we review research relevant to our study, starting with traditional, non-neural approaches to automated memory leak detection---covering both binary and source code analysis methods---and then surveying work that leverages large language models (LLMs) in source code analysis.


\subsection{Memory Leak Detection before LLMs}\label{ssec:related_memleak}

\subsubsection{Binary Analysis}\label{sec:binaryanalysis}

Binary analysis tools for memory management are essential for languages and environments where unused memory cleanup (garbage collection) might not happen automatically. The most popular memory analysis tool for C/C++ is \emph{Valgrind}~\cite{nethercote2007valgrind}, which can run the code in a ``controllable'' debugger-like environment and help identify memory leaks, uninitialized memory use, and invalid memory access among other issues. \emph{Valgrind}'s dynamic analysis is especially valuable in cases when the memory leak has already somehow revealed itself or for proactive memory testing during the development process. However, these scenarios are not in the focus of our study; our primary goal is ``scanning'' the whole codebase for potential problems. Other notable binary analyzers include \emph{Dr. Memory}\footnote{\url{https://github.com/DynamoRIO/drmemory}}~\cite{bruening2011practical}, 
\emph{PurifyPlus}\footnote{\url{https://www.unicomsi.com/products/purifyplus/}}, 
\emph{Insure++}\footnote{\url{https://www.parasoft.com/products/parasoft-insure/}}, 
\emph{Intel Inspector}\footnote{\url{https://www.intel.com/content/www/us/en/developer/tools/oneapi/inspector.html}}, and others.
%
Binary analysis can employ statistical methods such as PCA~\cite{li2020pca} and other machine learning techniques to uncover anomalies in memory access patterns~\cite{10.1145/3092566,10.1007/978-3-031-05237-8_66}.
\emph{AddressSanitizer}~\cite{serebryany2012addresssanitizer} (integrated into \emph{GCC} and \emph{CLang}; \emph{LeakSanitizer} is integrated into \emph{AddressSanitizer}) works at the level of source code and needs to be enabled during compilation. \emph{Electric Fence}\footnote{\url{https://packages.ubuntu.com/en/electric-fence}} represents another example of binary analysis; it hooks into memory allocation calls but does not annotate the binary itself; one could call it a runtime memory protection tool. Recently, attempts have been made to combine binary analysis with LLMs, e.g., the ChatWithBinary tool\footnote{\url{https://github.com/Protosec-Research/ChatWithBinary}}, but this field is still at its infancy.

\subsubsection{Source code Analysis}\label{sec:sourcecodeanalysis}

Tools from this class work directly with the project's code and do not require runtime execution of the programmed logic. These tools are more suitable for a broad scan of the project for memory-related problems, e.g. as a part of a standard continuous integration pipeline, and hence are more relevant for the present work. Important tools from this class include \emph{Infer}\footnote{\url{https://fbinfer.com/}}~\cite{10.1145/2049697.2049700,10.1007/978-3-642-20398-5_33,reynolds2002separation,calcagno2009compositional,calcagno2015open,blackshear2017open,o2019separation,distefano2019scaling}, including its \emph{Inferbo} library~\cite{yi2017inferbo} for buffer overrun analysis, and the line of research based on static value-flow analysis (SVF)~\cite{sui2016svf,10.1145/2338965.2336784,10.1145/2581122.2544154,10.5555/2190025.2190075,10.1145/2854038.2854043}. Industrial solutions for code-based analysis include \emph{Coverity}\footnote{\url{https://www.synopsys.com/software-integrity/static-analysis-tools-sast/coverity.html}}, SMOKE~\cite{fan2019smoke}, \emph{Sparrow}\footnote{\url{https://github.com/ropas/sparrow}}~\cite{jung2008practical}, 
KLEE~\cite{cadar2008klee}, Clang Static Analyzer\footnote{\url{https://clang-analyzer.llvm.org/}}~\cite{clang-analyzer}, and others.

\emph{K-Meld}~\cite{emamdoost2021detecting} proposes a Linux-kernel-scale approach to memory leak detection with a static kernel memory leak detector that constructs a whole-program representation (using LLVM IR and inter-procedural control-flow analysis) to automatically identify memory allocation functions and their corresponding deallocation functions via a usage- and structure-aware analysis coupled with context-aware rule mining. It employs an \emph{ownership reasoning} mechanism that uses enhanced escape analysis to determine when allocated objects escape (i.e., have their ownership transferred) and identifies consumer functions (callees that free or universally consume an object) to decide which code context is responsible for each memory release. K-Meld statically checks each allocation site's control flow for missing deallocation calls by pattern-matching common error-cleanup sequences. \emph{K-Meld} discovered $218$ new memory leak bugs ($41$ assigned CVEs) in the Linux kernel, with $115$ of these leaks found in specialized modules thus demonstrating its effectiveness in detecting leaks across even heavily customized kernel components.
{The code has been published by the authors}\footnote{{\url{https://github.com/Navidem/k-meld}}.}, but
appears to lack files needed to apply it to other codebases\footnote{{We were not able to run the \emph{K-Meld} tool or obtain these files from the authors.}}.

A more recent approach to memory leak detection called \emph{Memory Leak Hunter}~\cite{aslanyan2024combining} proposes to 
\begin{inparaenum}[(1)]
\item build a special ``ProcedureGraph'' structure combining the data flow information and the call graph, 
\item {annotate} the functions of the codebase via a limited number of rules into $7$ categories, and to exclude false positive detections via directed symbolic execution, using a version of \emph{KLEE}~\cite{cadar2008klee} modified for this purpose; the authors posit that this is the key step in false alarm removal.
\end{inparaenum}
Its main differences from our work are that Memory Leak Hunter relies on \emph{KLEE}'s dynamic symbolic execution (while we have focused on static 
{analysis}), is strictly formal (does not use any insights from natural language or prior code knowledge whatsoever), and is very task-specific. Moreover, some important implementation details (such as exact rules for annotating the methods) have not been published yet\footnote{
{The code is not publicly available as of early 2025.}}.

\subsection{LLMs for Source Code}\label{ssec:related_llm4code}

AI-powered approaches to source code analysis span several decades, and the same could be said about AI for applied software engineering~\cite{de1994artificial}. However, it is only with the introduction of Transformer-based~\cite{10.5555/3295222.3295349} contextualized word embeddings~\cite{devlin2019bert,feng2020codebert} and, later, large generative language models~\cite{radford2019language,brown2020language,achiam2023gpt} and ``AI assistants'' for software developers based on the latter that the field has gained unprecedented attention from the research community. Here we focus on the segment of this domain relevant to our study: research connecting LLMs and methods of formal code analysis.


%

\subsubsection{LLMs and Static Code Analysis}\label{ssec:related_llm_plus_static}

There is a line of research where LLMs and static code analysis methods work together, combining semantic code understanding of LLMs with exact knowledge provided by traditional methods. 
LLMs were used to enhance the traditional approach for the tasks of inferring error specifications in C code~\cite{chapman2024interleaving} and for code revision generation
\cite{wadhwa2024core}.
An interesting special case is the \emph{fault localization} task, where the objective is to detect the part of the code responsible for the bug~\cite{kang2024quantitative,yan2024better}. This type of code analysis has not been widely adopted by the industry so far because existing static analysis methods cannot provide helpful explanations of the root cause of the bug, and large language models have a potential to fix this drawback. 
LLMs have been successfully applied for automated program repair (APR), when the exact bug position is known, and the objective is to generate the bug fixing patch~\cite{hossain2024deep, zhou2024large}. 
At the same time, bug localization and vulnerability detection appear to be much harder for LLMs. Multiple recent studies demonstrate that LLM performance here is currently limited~\cite{anand2024criticalstudycodellmsdo,fang2024largelanguagemodelscode}, and novel research directions are required to improve LLM performance in this potentially very useful application~\cite{gao2023far, sun2024llm4vuln, ding2024vulnerability}.
Static analysis tools improve the above solutions providing structured code information to the LLM context, which is especially important in the case of a large codebase~\cite{chen2024large, zhang2024llms, wang2023defecthunter}. In~\cite{10.1145/3510003.3510153}, LLM were used to filter false warnings of static analysis tools.
Recently, LLM-based agentic tools gained popularity for complex code analysis tasks due to their multi-strategy nature~\cite{10.1145/3712003}. GPT-4 was applied to emulate pseudocode execution for fault localization through static analysis steps such as code retrieval and call graph reconstruction~\cite{DBLP:journals/corr/abs-2312-08477}. Skip-analyzer~\cite{DBLP:journals/corr/abs-2310-18532}, a two-stage agentic ChatGPT-based tool, detects bugs like resource leaks and null pointer dereferences, filtering out false positives. Its resource leak dataset contains 46 instances collected by the Infer analyzer, demonstrating GPT-4's efficiency for resource leak detection. Similarly, the work~\cite{DBLP:journals/corr/abs-2311-04448} employs LLM prompts to locate resource acquisition and release, then analyzing execution paths between these points.
In contrast, we propose combining LLMs with existing static analyzers during a function-level annotation stage, specifically targeting memory leaks. Our approach demonstrates effectiveness on both extensive synthetic datasets and real-world open-source projects.
%
A few works have introduced combinations of LLMs with {code analysis}
engines, using LLMs to locate critical sections in code, reduce the symbolic engine load, or infer conclusions about undecided paths~\cite{li2024enhancing}; in~\citet{liu2024propertygpt}, LLM generates function properties descriptions for further processing by symbolic engine,
{which is then applied}
to smart contract verification.


\section{Conclusion}\label{sec:conclusion}


In this work, we have presented LAMeD, a novel approach to automatically generating annotations for functions in source code that we have shown to be helpful for memory leak detection.
Our experiments show that LLM-generated annotations are very precise (\emph{Codestral} on manually labeled \texttt{cJSON} has precision $0.933$); however, the recall score is lower, on average only approaching $0.6$, which means that more functions could be annotated as sources or sinks.

Moreover, generated annotations improve the memory leak detection accuracy for both CodeQL and Cooddy analyzers in almost all projects.
The total number of warnings generated by both analyzers has notably increased after using LAMeD annotations.
Note, however, that such a boost was observed not for all projects.
%
%
%

We believe that we have only scratched the surface of what can be done in this approach. First, it appears potentially beneficial to generate meaningful annotations for functions that are called yet not implemented in the codebase, in particular items imported from external libraries. 
Second, as a step particularly useful for the software engineering community, one could execute LLM-based generation of memory-related annotations for functions in the most widely used open source libraries (such as, e.g., \emph{openssl}). Thus, prepared (and then manually validated) configuration files would provide an analyzer with much more information for reasoning and inevitably allow it to reduce the percentage of false positive detections: analyzers commonly ``expect the worst'' from calls of external items and hence produce a large number of false alarms.

As future work, in addition to memory leak detection, we posit that LLMs could be employed to detect other critical software bugs, such as buffer overflows or race conditions. Our preliminary experiments (not reported in this work) confirm the feasibility of this approach, which could extend the utility of LLMs in static analysis across a broader range of defects, significantly improving software reliability. Additionally, apart from bugs and vulnerabilities detectable by available tools such as  \emph{Infer}, \emph{CodeQL}, \emph{Coverity}, \emph{Cooddy}, and others, one could consider using a similar LLM-enriched approach to the errors for which formal analysis techniques without human work might arguably be insufficient.

\section{Data Availability}\label{sec:availability}

We are releasing supplementary materials needed to replicate the results of this work, including 
{fully manually annotated functions of the open source \emph{cJSON} library, }
{real-life memory leaks collected from GitHub for analyzers' capability evaluation}, and prompts we have used for LLMs, anonymously on Zenodo\footnote{\url{https://zenodo.org/communities/llm-4-lsr}}. We plan to make an official open release upon acceptance.

\bibliographystyle{ACM-Reference-Format}
\bibliography{99_references}
\end{document}